\documentclass[11pt]{article}
\usepackage{epsfig}
\usepackage{cite}
\title{\bf Reggeon-gluon vertices with Ward identities}
\author{M.A.Braun, M.I.Vyazovsky\\
{\it S.Petersburg State University, Russia}}

\begin{document}

\maketitle
\input epsf

\def\beq{\begin{equation}}
\def\eeq{\end{equation}}
\def\lra{\leftrightarrow}
%%%%%%%%%%%%%%%%%%%%%%%%%%%%%
\def\pd{\partial}
\def\ab{{\alpha_2}}
\def\aa{{\alpha_1}}
\def\ba{{\beta_1}}	
\def\np{n^+}
\def\nm{n^-}
\def\tr{{\rm Tr}\,}

{\bf Abstract}

Ward identities for reggeons are studied in the framework of effective action approach
to the QCD in Regge kinematics. It is shown that they require introduction of new contributions
not present in the reggeon diagrams initially. Application to vertices RR$\to$RP and
RR$\to$RRP are considered and diagrams which have to be added to the QCD ones are found.

\vskip 0.3cm

\section{Introduction}
Strong interactions at high energies in the framework of the
perturbative QCD in Regge kinematics can be described in terms of
reggeized gluons (''reggeons''), which combine into colourless pomerons
exchanged between colliding hadrons. Reggeons and their interactions
were first introduced in the dispersion approach using multiple unitarity cuts
~\cite{bfkl1, bfkl2, bartels1, bartels2}. Later, mostly to describe next order contributions,
a convenient and powerful method of effective action was proposed
in which the reggeons figure as independent dynamical fields interacting with
the standard gluons ~\cite{lipatov0,lipatov}. The effective action allows to present
scattering amplitudes in the Regge kinematics as a sum of diagrams, similar to the
Feynman ones with certain rules for propagators and interaction vertices
~\cite{antonov}. The latter, apart from the standard QCD vertices, include the so-called induced
vertices in which the reggeons interact with two or more gluons.
With the growth of the number of participants the number of
such diagrams grows and their form becomes more complicated. In ~\cite{bpsv15} the vertex
for emission of a gluon in interaction of two reggeons was calculated. It contained quite
a number of terms. In the light-cone variables
separate terms, as function of the longitudinal variables, were found to fall not fast
enough  to perform subsequent integration necessary for construction
of physical amplitude, although the total vertex
had a good property in this respect.

In search for a method to improve convergence in ~\cite{BLV}
Ward identities for reggeon amplitudes were proposed, which later were used to
calculate the next order kernel for the interaction of three reggeons in the odderon
configuration ~\cite{BFLV}. These Ward identities were based on the comparison of the diagrams
including a reggeon with the ones in which this reggeon is replaced by the gluon.
Obviously, the latter diagrams should not contains the ones with induced vertices for
this particular reggeon. Thus the set of diagrams used for the Ward identities turned out to be
narrower than the initial set of reggeon diagrams. Use of  Ward identities in ~\cite{BFLV}
allowed to exclude nearly all diagrams with induced vertices and achieve better convergence
for the remaining ones.
A brief review of the effective action and Ward identities introduced in ~\cite{BLV} is presented in the
next section.

In this note we study this problem in a more general context, having in mind colour configurations
different from the simple odderon one. We show that generally not only the diagrams with the induced
vertices for a particular reggeons are to be excluded from the Ward identity for it, but also some
new diagrams have to be added, which did not exist initially in the reggeon perturbation theory.
It turns out that in this general case the use of Ward identities, apart from the diagrams
with the QCD vertices, a whole set of special additional diagrams has to be included to correctly
describe the amplitudes. In the highly symmetric odderon configuration most of these additional
diagrams indeed cancel and the simple result of ~\cite{BFLV} follows in the particular gauge used there.
However, in more general configurations or gauges the additional diagrams are found to be much more
numerous and complicated, which puts the advantage of using Ward identities in some doubt.

We start our study in Section 3 with a simple case of the vertex RR$\to$RP for fusion
of two reggeons into one with emission of
a real gluon. This vertex served as an intermediate tool for the derivation of the second order
odderon interaction in ~\cite{BFLV}. So we compare this symmetric case with our general one
in Section 4. In Section 5 we consider a more complicated vertex RR$\to$RRP
for emission of a gluon in interaction of two reggeons to see how our results change with the number
of participating reggeons. The results of this section together with those in Section 3 clearly
show the general pattern valid for arbitrary number of reggeons.

\section{Effective action and Ward identities for the vertices}
The effective action formalism considers interaction at a given rapidity.
Apart from the normal gluon (''particle'') field $v$  an independent reggeon field $A$ is introduced,
which interacts with gluon via the so-called induced vertices and connects particles with large
rapidity distance.
The Lagrangian (at given rapidity) is
\cite{lipatov0,lipatov}:
\beq
{\cal L}_{eff}={\cal L}_{QCD}(v)
+ 2 \tr\Big\{{\cal V}_+(v_+)\pd^2_{\perp}A_-
+ {\cal V}_-(v_-)\pd^2_{\perp} A_+\Big\}.
\label{e1}
\eeq
Here ${\cal L}_{QCD}(v)$ is the standard QCD Lagrangian for the gluon field.
In the second, ''induced'', term  $a_{\pm}$ denote $\pm$ components of a 4-vector $a$ in the light-cone
metric: $ a_\pm=(a_0\pm a_3)/\sqrt{2}$ and
$$
{\cal V}_{\pm}(v_{\pm})=-\frac{1}{g}\pd_{\pm}\frac{1}{D_{\pm}}\pd_{\pm}*1=
\sum_{n=0}^{\infty}(-g)^nv_{\pm}(\pd_\pm^{-1}v_\pm)^n
$$
\beq
=v_{\pm}-gv_{\pm}\pd_\pm^{-1}v_\pm+g^2 v_{\pm}\pd_\pm^{-1}v_\pm
\pd_\pm^{-1}v_\pm -... \ .
\label{e2}
\eeq
The reggeon fields $A_+$ and $A_-$ describe incoming and outgoing reggeons
respectively. In accordance with Regge kinematics they satisfy
\
\beq
\partial_- A_+ =
\partial_+ A_- =0 .
\label{e5}
\eeq
In the momentum space these conditions transform into the requirement that the
momentum of the incoming (outgoing) reggeon has its minus (plus) component equal to zero.

In perturbation theory the effective Lagrangian (\ref{e1}) generates diagrams, which apart
from the standard QCD ones include those with induced vertices describing interaction of
a reggeon with one or several gluons according to (\ref{e2}). The rules for constructing these
vertices and an explicit form of some of the lowest ones was presented in \cite{antonov}.
The simplest of them is the vertex $V_0$ for transition of a reggeon into the gluon
$A_{\pm}^a\to v_\mu^b$
\beq
V_0=iQ^2n^{\mp}_\mu \delta^{ab} \ ,
\label{v0}
\eeq
where $a$ and $b$ are colour indices, $Q$ is the reggeon momentum
(obviously conserved in the transition) and $n^{\mp}$ are pure longitudinal
unit vectors: $n^-_+=n^+_-=1$, $n^{\pm}_\perp=0$. Vertices for transition into more
gluons necessary for our results will be presented later.

Note that, due to the form of $V_0$, transition of a reggeon into the gluon
makes the diagram identical with the one in which the reggeon is replaced by the gluon.
The only difference is that the reggeon polarization vector is $n^-$ for the incoming reggeon and
$n^+$ for the outgoing one and for the real gluon we have to use normal polarization vectors $e$,
which also contain transverse components. This observation was the basis of the derivation of
the Ward identities for reggeon vertices.
In the general case, the amplitude is given by product
of the vertex $\Gamma$ for interaction of $k$ gluons and $l$ reggeons
(incoming and outgoing) with their appropriate polarization vectors:
$$
{\cal A} = e^1_{\mu_1} \cdots e^k_{\mu_k}
\cdot n^1_{\alpha_1} \cdots n^l_{\alpha_l}
\Gamma_{\alpha_1 \dots \alpha_l \mu_1 \dots \mu_k} \ .
$$
Now consider
the standard Ward identity
for the vertex $\bar\Gamma$ for interaction of $k+m$ gluons and $l-m$ reggeons,
where polarization vectors of $m$ gluons are replaced by their momenta $Q_i$
which are supposed equal to momenta of reggeons $1...m$ in $\Gamma$:
\begin{equation}
e^1_{\mu_1} \cdots e^{k}_{\mu_{k}}
\cdot (Q_1)_{\alpha_1} \cdots (Q_m)_{\alpha_m} \cdot
n^{m+1}_{\alpha_{m+1}} \cdots n^l_{\alpha_l}
\bar\Gamma_{\alpha_1 \dots \alpha_l \mu_1 \dots \mu_k}
=0 .
\label{ew1}
\end{equation}
In ~\cite{BLV} it was noted that if one drops
all diagrams for $\Gamma$ in which reggeons $1...m$ interact with two,
three and more gluons and only the simple transition into a single gluon is left,
then the remaining set of diagrams will be identical to the one for $\bar\Gamma$.
So the Ward identity (\ref{ew1}) becomes a relation between
a certain subset of diagrams of the amplitude $\cal A$. 

In ~\cite{BLV} it was shown that application of these Ward identities allows, first, to improve convergence
of the effective action diagrams at large longitudinal momenta and, second, may hopefully eliminate at least some
of the diagrams with induced vertices altogether. This hope was later confirmed in the calculation of
the next order corrections to the odderon kernel in ~\cite{BFLV}, where all but one induced diagrams
for the vertex RR$\to$RP were eliminated after multiple application of the Ward identities.
In the next section we shall demonstrate that this result in fact refers to the particular configuration
and chosen gauge in the study of ~\cite{BFLV}. In a more general configuration and gauge
the reduction in the number of diagrams with induced vertices is more modest.

\section{Vertex RR$\to$RP}
\subsection{Notations and diagrams}
We study the amplitude for gluon production ${\cal A}$ in the fusion of two
reggeons into one
\beq
R_{a_2}(Q_2)+R_{a_1}(Q_1)\to R_{b}(R)+g_c(p),
\eeq
where $a_2,a_1,b$ and $c$ are colour indices.
This amplitude is a product of vertex $\Gamma$ for the transition RR$\to$ RP
with the gluon polarization vector $e$
\beq
{\cal A}=(e\Gamma)=e_\mu \Gamma_\mu.
\eeq
In the following multiplication by $e$ will in many cases not be written explicitly.
We use the standard light-cone metric and only covariant components. Summation over repeated
4-vector indices is assumed in the Lorentz light-cone metric.
The amplitude splits into two independent parts with colour indices
$C_1=f^{a_2cd}f^{ba_1d}$ and $C_2=f^{a_1cd}f^{ba_2d}$. We study the part with the colour vertex
$C_1$ here. The other part can be studied in quite the similar manner.

\begin{figure}
\hspace*{2 cm}
\epsfig{file=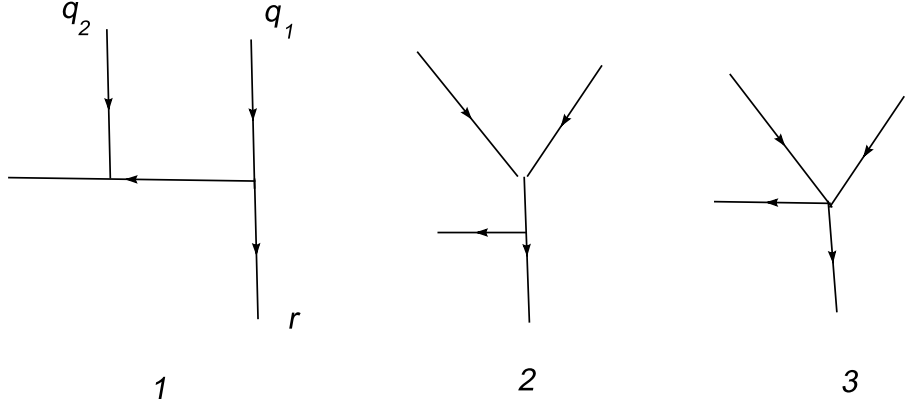, width=12 cm}
\caption{Diagrams for the vertex RR$\to$RP with QCD vertices.
Simple external lines directed upwards and downwards actually refer to reggeons.}
\label{fig1}
\end{figure}

\begin{figure}
\hspace*{2 cm}
\epsfig{file=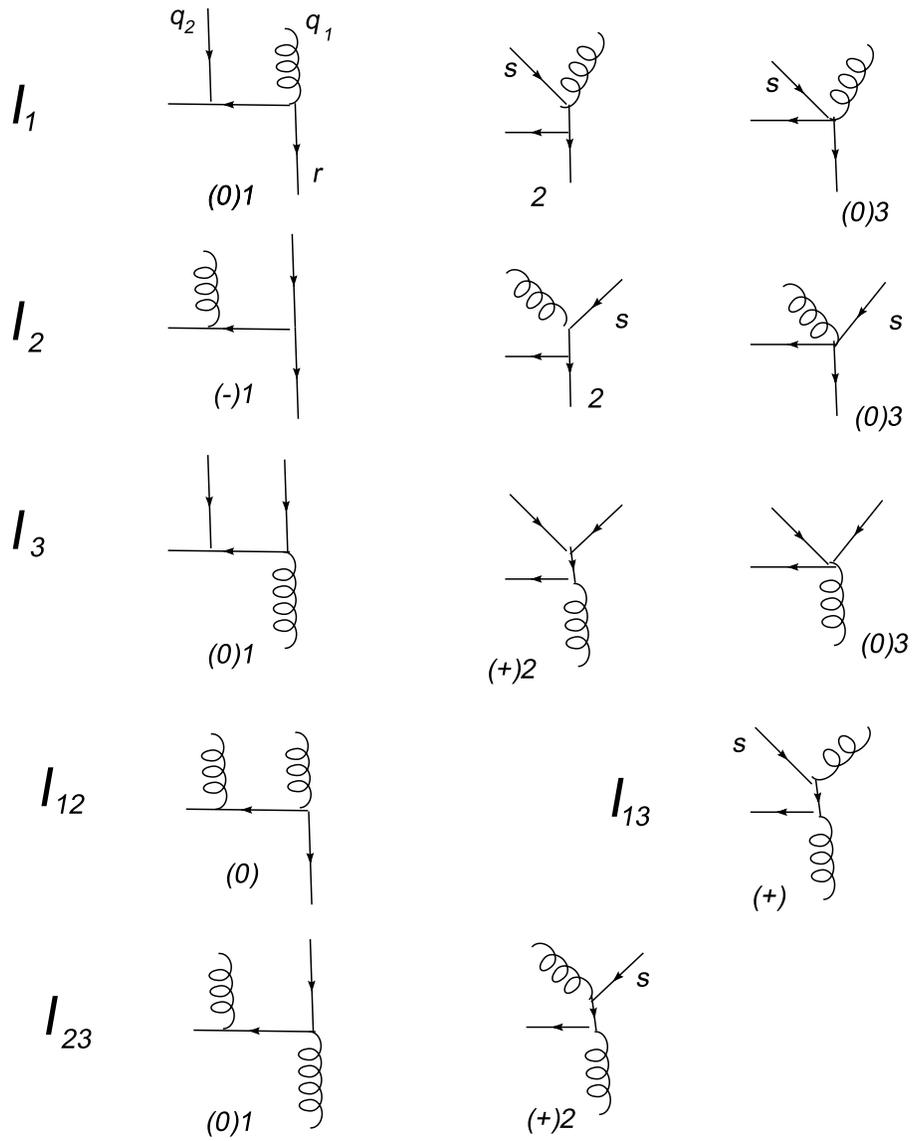, width=12 cm}
\caption{Diagrams for the vertex RR$\to$RP and Ward identities with induced vertices.
Simple external lines directed upwards and downwards actually refer to reggeons.}
\label{fig2}
\end{figure}

As explained in the previous section, we introduce polarization vectors for reggeons: $n^-$
for the incoming reggeons with momenta
$Q_1$ and $Q_2$ and $n^+$ for the outgoing reggeon with momentum $R$. To simplify notation
in the following we denote the outgoing reggeon by index 3, so that $Q_3\equiv R$.
Then  vertex RR$\to$RP will be given by the convolution
\beq
\Gamma_\mu=(n_1n_2n_3 D)_\mu\equiv n^-_{\alpha_1}n^-_{\alpha_2}n^+_\beta D_{\alpha_1\alpha_2\beta\mu},
\label{vert1}
\eeq
where $D$ is the sum of 9 diagrams shown in Figs. \ref{fig1} and \ref{fig2}.
In these diagrams it is assumed that simple external lines directed upwards and downwards actually refer
to reggeons, which go into gluons via vertex $V_0$ (Eq. (\ref{v0})).
They include  purely QCD diagrams $D_0$ (Fig. \ref{fig1}) and part of the diagrams
with induced vertices for reggeons 1,2 and 3 (Fig. \ref{fig2})

\beq
D=D_0+I_{1,1}+I_{2,1}+I_3+I_{23,1}
\label{d1}
\eeq
(9 diagrams in all).
Here $I_i$ is the sum of diagrams with only a single induced vertex
for reggeon $i$, $I_{i,k}$ is the $k$th diagram for $I_i$. Likewise $I_{ik}$ is the sum of diagrams with
two induced vertices for reggeons $i$ and $k$ and $I_{ik,l}$ is the $l$th diagram for this sum
in Fig. \ref{fig2}.

\begin{figure}
\hspace*{2 cm}
\epsfig{file=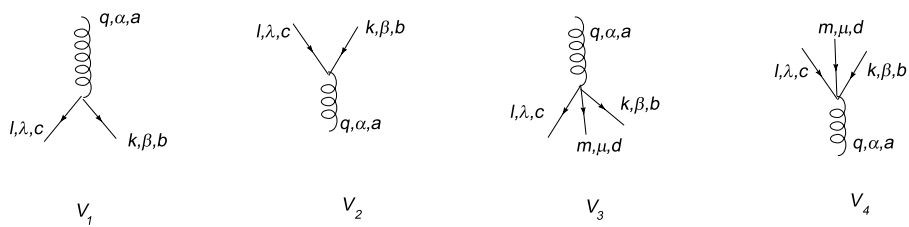, width=12 cm}
\caption{Induced vertices for transition of the reggeon into gluons}
\label{fig3}
\end{figure}

Since the induced vertex for transition of the reggeon into gluon is trivial, a simple external
line in Figs. \ref{fig1} and \ref{fig2} is just unity and vertices connecting simple lines are the standard QCD vertices.
As to the vertices for transition of the reggeon into 2 or 3 gluons, they are shown in Fig. \ref{fig3} and given by
(see ~\cite{antonov})
\[V_1= gq^2f^{acb}\frac{1}{l_-}n^-_\lambda n^-_\beta n^+_\alpha\ ,\]
\[V_2= -gq^2f^{acb}\frac{1}{l_+}n^+_\lambda n^+_\beta n^-_\alpha\ ,\]
\[V_3= ig^2 q^2n^-_\lambda n^-_\mu n^-_\beta n^+_\alpha \Big(\frac{f^{bce}f^{dea}}{m_-l_-}+
\frac{f^{dce}f^{bea}}{k_-l_-}\Big),\]
\beq
V_4= ig^2 q^2n^+_\lambda n^+_\mu n^+_\beta n^-_\alpha \Big(\frac{f^{bce}f^{dea}}{m_+l_+}+
\frac{f^{dce}f^{bea}}{k_+l_+}\Big).
\label{rules}
\eeq
The meaning of indices $s$, $(0)$, $(+)$ and $(-)$ will be explained later.

We are going to use the Ward identities introduced in ~\cite{BLV} for the
calculation of the amplitude ${\cal A}$. As explained in the previous section, they are based on
the comparison of
the reggeon diagrams with the
ones where a particular reggeon is replaced by the gluon. In such diagrams this particular
reggeon obviously cannot be coupled with induced vertices with more than one gluon.
So these Ward identities actually
should be applied only to a part of all contributions which does not contain higher induced vertices for
the considered reggeon. However, this is not the end of the story. In fact, as we shall see,
the amplitude with the reggeon
replaced by the gluon may contain additional diagrams, which are absent for the reggeon.
Such diagrams do not give contribution for the amplitude ${\cal A}$ but do give contributions to the
Ward identities.

In our case these additional diagrams are diagrams in
Fig. \ref{fig2} with a single induced vertex $I_{i,2}$ and
$I_{i,3}$ where $i=1,2$ and with two induced vertices $I_{13}$ and $I_{23,2}$.

One observes that some of the diagrams with induced vertices may in fact give zero
for the vertex $\Gamma$
after multiplication by the product $n^- n^- n^+$ in (\ref{vert1}). In our case
among the initial reggeon diagrams  $I_{3,2}$ vanishes in (\ref{vert1})
due to  the property of the three-gluon vertex in the Regge kinematics.

Also the diagrams in Fig. \ref{fig2} added to the initial ones  to cover application of Ward identities
vanish in (\ref{vert1}) for the same reason. However, one should take certain care with these diagrams.

It is instructive to study their explicit form.
As an example, take $I_{1,2}$ with a single induced vertex for reggeon 1. It is given by
\beq
I_{1,2}=-n^+_{\alpha_1}n^-_{\alpha_2}n^-_\nu
\frac{Q_1^2}{Q_{2-}(Q_1+Q_2)^2}\gamma_{\nu\beta\mu}(Q_1+Q_2,-R),
\eeq
where $\gamma_{\mu\nu\sigma}(Q,R)$ is the three-gluon vertex for incoming
momenta $Q_\mu,R_\nu$ and $-(Q+R)_\sigma$. Remarkably $I_{1,2}$ contains $Q_{2-}$
in the denominator and has no sense when gluon 2 becomes the reggeon with $Q_{2-}=0$.
So in fact  multiplied by the reggeon 2 polarization vector
$n^-_{\alpha_2}$, it contains an indefinite factor $(n^-)^2/Q_{2-}$.
To be able to include such a diagram into the general form (\ref{vert1}) and (\ref{d1})
we have to impose the condition
\[ n^-_{\alpha_2}[I_{1,2}]_{\alpha_1\alpha_2\beta\mu}=0 \]
equivalent to the interpretation
\beq
\frac{{n^-}^2}{Q_{2-}}\Big|_{Q_{2-}=0}=0.
\label{cond}
\eeq
Then this diagram will not appear in the expression for (\ref{vert1}) for $\Gamma$
and we can include it as a singular contribution into $I_1$.
However, it gives a non-zero contribution to the Ward identity for reggeon 2.
Indeed, multiplication by scaled $Q_2/Q_{2+}$ replaces $Q_{2-}$ by $Q_{2+}$ in the denominator
and makes it perfectly valid for the Regge kinematics with $Q_{2-}=0$.
The same is true for all the rest of similar diagrams in Fig. \ref{fig2}. To indicate
appearance of such singular denominators we mark the corresponding external legs by
letter $s$ in Fig. \ref{fig2}.

So we conclude that to apply Ward identities it is not sufficient to drop the diagrams
with transition of a chosen reggeon into two or more gluons. One has also to add some new
diagrams, which are absent for the diagram with the chosen reggeon but are present when it is
replaced by the gluon.

\subsection{Ward identities}
In our derivation to simplify notations we normalize the
incoming reggeon momenta putting
\[ \frac{Q_i}{Q_{i+}}= q_i\ ,\ \ i=1,2\]
and for the outgoing reggeon
\[ \frac{R}{R_-}\to q_3.\]
So we denote momenta of all reggeons, incoming and outgoing,
by just $q_1,q_2,q_3$ and their polarization vectors as
$n_1,n_2,n_3$ having in mind that for the incoming reggeons we have to take
$n^-$ and for the outgoing reggeons $n^+$. We denote the transverse
momenta of all reggeons, duly normalized,  by $t_1,t_2,t_3$, so that
in these notations
\beq
n=q-t.
\label{n}
\eeq

Using (\ref{n}) we rewrite (\ref{vert1}) as
\beq
\Gamma=(q_1-t_1)(q_2-t_2)(q_3-t_3) D .
\label{vert2}
\eeq
We find 8 terms. Taking into account that induced vertices are longitudinal
we find these terms as
\[a_1=q_1q_2q_3 D,\]
\[a_2=-q_1q_2t_3(D_0+I_1+I_2+I_{12}),\]
\[a_3=-q_1t_2q_3(D_0+I_1+I_3+I_{13}),\]
\[a_4=-t_1q_2q_3(D_0+I_2+I_3+I_{23}),\]
\[a_5=q_1t_2t_3(D_0+I_1),\]
\[a_6=t_1t_2q_3(D_0+I_3),\]
\[a_7=t_1q_2t_3(D_0+I_2),\]
\[a_8=-t_1t_2t_3D_0\ .\]

We want to calculate these term using Ward identities (WI)
\beq
q_1n_2n_3(D_0+I_2+I_3+I_{23})=0 ,
\label{wi1}
\eeq
\beq
q_1q_2n_3(D_0+I_3)=0 ,
\label{wi2}
\eeq
4 similar ones obtained after permutations
(123)$\to$(231) and (123)$\to$(312) and
\beq
q_1q_2q_3D_0=0 .
\label{wi3}
\eeq

Our strategy will be to express from WI the result of
contraction of momenta $q_i$ with $D_0$.
Take (\ref{wi1}). It reads
\beq
q_1(q_2-t_2)(q_3-t_3)(D_0+I_2+I_3+I_{23})=0
\label{wi11}
\eeq
or
\beq
q_1t_2t_3D_0=q_1t_2q_3(D_0+I_3)+q_1q_2t_3(D_0+I_2)-q_1q_2q_3(I_2+I_3+I_{23}) ,
\label{wi12}
\eeq
where we used (\ref{wi3}).
We put it into $a_5$
\beq
a_5=q_1t_2t_3I_1+q_1t_2q_3I_3+q_1q_2t_3I_2-q_1q_2q_3(I_2+I_3+I_{23})+
(q_1t_2q_3+q_1q_2t_3)D_0\ .
\label{t5}
\eeq

Now we use (\ref{wi2})
\[
q_1q_2(q_3-t_3)(D_0+I_3)=0 ,
\]
which with (\ref{wi3}) taken into account gives
\beq
q_1q_2t_3D_0=q_1q_2q_3I_3 \ .
\label{wi21}
\eeq
Smilarly,
\beq
q_1t_2q_3D_0=q_1q_2q_3I_2 \ .
\label{wi22}
\eeq
Put into (\ref{t5}) it changes the last term into
$q_1q_2q_3(I_2+I_3)$ and cancels two first term in the
preceding term.
We get
\beq
a_5=q_1t_2t_3I_1+q_1t_2q_3I_3+q_1q_2t_3I_2-q_1q_2q_3I_{23}\ .
\label{t51}
\eeq
Similarly,
\beq
a_6=q_3t_2t_1I_3+q_3t_2q_1I_1+q_3q_2t_1I_2-q_1q_2q_3I_{12}\ ,
\label{t6}
\eeq
\beq
a_7=q_2t_1t_3I_2+q_2t_1q_3I_3+q_2q_1t_3I_1-q_1q_2q_3I_{13}\ .
\label{t7}
\eeq

Now take $a_2$ and use (\ref{wi21})
\[a_2=-q_1q_2t_3(I_1+I_2+I_{12})-q_1q_2q_3I_3\ .\]
Similarly,
\[a_3=-q_1t_2q_3(I_1+I_3+I_{13})-q_1q_2q_3I_2\ ,\]
\[a_4=-t_1q_2q_3(I_2+I_3+I_{23})-q_1q_2q_3I_1\ .\]

In the sum
\[\sum _{i=2}^7a_i
=q_1t_2t_3I_1+q_3t_2t_1I_3+q_2t_1t_3I_2
-q_1q_2t_3I_{12}
-q_1t_2q_3I_{13}
-t_1q_2q_3I_{23}\]\[
-q_1q_2q_3(I_1+I_2+I_3+I_{12}+I_{13}+I_{23}).\]
The last term cancels $a_1$ and we finally get
\beq
\Gamma_{RR\to RP}=
-t_1t_2t_3D_0+q_1t_2t_3I_1+t_1q_2t_3I_2+t_1t_2q_3I_3
-q_1q_2t_3I_{12}
-q_1t_2q_3I_{13}
-t_1q_2q_3I_{23} \ .
\label{vert21}
\eeq

Here factors $q_i$ applied to reggeons from induced vertices
can be replaced by $n_i$, since reggeon legs are
longitudinal.

At this moment it is necessary to take into account condition (\ref{cond}) for the singular
parts of contributions from Fig. \ref{fig2} with index $s$ at some external legs.
Condition (\ref{cond}) allows to substitute in such diagrams $t\to q$ for this leg, which cancels
the singularity due to the denominator.

%%%%%%%%%%%%%%%%%%%%%%%%%%%%%%%%%%%%%%%%%%%%
\section{Additional terms in ${\cal A}$}
The attractive feature of application of Ward identities to the calculation
of high energy amplitudes represented by reggeon diagrams is that in  the framework of
effective action it allows, at least partially,  to exclude contributions from induced vertices
leaving only QCD diagrams convoluted with transverse momenta. In our case such QCD contribution
is given by term $-t_1t_2t_3D_0$. However, Eq. (\ref{vert21}) shows that in fact
to the QCD contribution certain terms have to be added, which include diagrams with induced vertices
and moreover some diagrams absent at all in the reggeon diagram technique. In this section we study this
additional contributions in Eq.  (\ref{vert21}). As we shall find many of the additional terms
are in fact zero. However not all, so that the QCD term alone cannot give the correct value
for the amplitude
(on mass shell, multiplied by the polarization vector $e$).

To locate diagrams which actually vanish we note that all external lines which eventually
transform into reggeons
and do not bear index $s$ are to be multiplied by the appropriate transverse vector $t$.
They can be attached either to QCD vertices or to induced ones. In the latter case multiplication
by the transverse vector $t$ will give zero, as is evident from rules (\ref{rules}).  This removes
the contribution from all diagrams in Fig. \ref{fig2} marked by index $(0)$. Then we are left with
only 6 diagrams $I_{1,2}$, $I_{2,1}$, $I_{2,2}$, $I_{3,2}$, $I_{13}$ and $I_{23,2}$.

Some further reduction of the number of contributing diagrams can be achieved choosing special gauges
with either $e_+=0$ or $e_-=0$. In fact the line corresponding to the emitted gluon can also be
attached either to the QCD vertex or to the induced one. In the latter case it bears factor $n^\pm$
depending on which induced vertex it is attached to. In Fig. \ref{fig2} such diagram is marked
by index $(\pm)$. It is clear that all diagrams marked by $(+)$ will vanish in the gauge
with $e_+=0$ and all diagrams marked by $(-)$ will vanish in the gauge $e_-=0$. In particular,
in the commonly assumed gauge $e_+=0$ the only remaining additional diagrams
are $I_{1,2}$, $I_{2,1}$ and $I_{2,2}$.

In the following for illustration we calculate all additional contributions which are non-zero in
the general gauge.

{\bf Term $t_1n_2t_3I_{2,1}$}%$

The diagram  is given by
\beq
I_{2,1}=-n^+_{\alpha_2}n^-_\nu n^-_\mu\frac{Q_2^2}{T^2 R_-}\gamma_{\alpha_1\beta\nu}(Q_1,-R)
\eeq
with $T=Q_1 -R$.
We find
\beq
t_1n_2t_3I_2=-n^-_\mu t_{1\alpha_1}t_{3\beta}n^-_\nu\gamma_{\alpha_1\beta\nu}(Q_1,-R)\frac{Q_2^2}{T^2 R_-}
\label{add1}
\eeq
and the contribution to the amplitude ${\cal A}$ will be
\beq
\Delta_{2,1}=-e_- t_{1\alpha_1}t_{3\beta}n^-_\nu\gamma_{\alpha_1\beta\nu}(Q_1,-R)\frac{Q_2^2}{T^2 R_-}.
\label{del5}
\eeq
So this term is given by the product of the QCD vertex on the right
convoluted with transverse momenta of reggeons 1 and 3 and multiplied by the
induced vertex for reggeon 2 on the left. It is the term which was explicitly introduced
in the calculations of the transition RRR$\to$RRR in the odderon case in ~\cite{BFLV}.

{\bf Term $t_1t_2q_3I_{3,2}$}%$

We have
\[I_{3,2}=-n^+_\nu n^-_\beta n^+_\mu\frac{R^2}{p_+\tilde{T}^2}\gamma_{\alpha_1\nu\alpha_2}(Q_1,-\tilde{T})\]
with $\tilde{T}=Q_1+Q_2$.
So
\[t_1t_2n_3 I_{3,2}=-n^+_\mu\frac{R^2}{p_+\tilde{T}^2}n^+_\nu t_{1\alpha_1}t_{\alpha_2}
\gamma_{\alpha_1\nu\alpha_2}(Q_1,-\tilde{T}).\]
One has
\[
n^+_\nu\gamma_{\alpha_1\nu\alpha_2}(Q_1,-\tilde{T})=
(Q_2-Q_1)_+g_{\alpha_2\alpha_1}+(2Q_1+Q_2)_{\alpha_2}n^+_{\alpha_1}-(2Q_2+Q_1)_{\alpha_1}n^+_{\alpha_2} .
\]
As a consequence
\beq
t_1t_2n_3 I_{3,2}=-n^+_\mu(Q_2-Q_1)_+\frac{R^2(t_1t_2)}{p_+\tilde{T}^2} .
\label{ad7}
\eeq
This leads to the additional contribution to the amplitude
\beq
\Delta_{3,2}=-e_+(Q_2-Q_1)_+\frac{R^2(t_1t_2)}{p_+\tilde{T}^2} .
\label{del7}
\eeq
So we obtain a non-zero additional contribution.
Like $\Delta_{2,1}$ it is given by the product of the QCD vertex on the top
convoluted with transverse momenta of reggeons 1 and 2 and multiplied by the
induced vertex for reggeon 3 on the bottom.
However, it vanishes in the gauge with $e_+=0$. It also vanishes for the sum
of contributions with the order of $Q_1$ and $Q_2$ reversed (''after symmetrization'').
This is why it did not appear in ~\cite{BFLV} where such symmetrization occurred due to
the odderon colour structure.

{\bf Term $n_1t_2t_3I_{1,2}$}%$

The explicit expression for $I_{1,2}$ is
\[I_{1,2}=-n^+_{\alpha_1}n^-_\nu n^-_{\alpha_2}\frac{Q_1^2}{Q_{2-}\tilde{T}^2}\gamma_{\nu\beta\mu}(\tilde{T},-R).\]
As explained one should substitute factor $q_2$ for $t_2$ which cancels the denominator $Q_{2-}$.
Then
\[q_2 I_{1,2}=-n^+_{\alpha_1}n^-_\nu\frac{Q_1^2}{Q_{2+}\tilde{T}^2}\gamma_{\nu\beta\mu}(\tilde{T},-R),\]
so that
\[n_1q_2t_3I_{1,2}=-t_{3\beta}n^-_\nu\frac{Q_1^2}{Q_{2+}\tilde{T}^2}\gamma_{\nu\beta\mu}(\tilde{T},-R).\]
We have
\[\gamma_{\nu\beta\mu}(\tilde{T},-R)=2R_{\mu}g_{\beta\nu}+(p-R)_\nu g_{\beta\mu}-(p+\tilde{T})_\beta
g_{\mu\nu}\ ,\]
so that
\beq
n_1q_2t_3I_{1,2}=\Big(2R_-t_{3\mu}+(t_3,p+\tilde{T})n^-_\mu\Big)\frac{Q_1^2}{Q_{2+}\tilde{T}^2}
\eeq
and the additional term in the amplitude ${\cal A}$ will be
\beq
\Delta_{1,2}=\Big(2(eR)_\perp+e_-\frac{(R,p+\tilde{T})_\perp}{R_-}\Big)\frac{Q_1^2}{Q_{2+}\tilde{T}^2}.
\label{del11}
\eeq

{\bf Term $t_1n_2t_3I_{2,2}$}%$

The explicit expression for $I_{2,2}$ is
\[I_{2,2}=n^+_{\alpha_2}n^-_\nu n^-_{\alpha_1}\frac{Q_2^2}{Q_{1-}\tilde{T}^2}\gamma_{\nu\beta\mu}(\tilde{T},-R).\]
We change $t_1\to q_1$, which gives
\[q_1I_{2,2}= n^+_{\alpha_2}n^-_\nu\frac{Q_2^2}{Q_{1+}\tilde{T}^2}\gamma_{\nu\beta\mu}(\tilde{T},-R),\]
so that
\beq
q_1n_2t_3I_{2,2}=\Big(-2R_-t_{3\mu}-(t_3,p+\tilde{T})n^-_\mu\Big)\frac{Q_2^2}{Q_{1+}\tilde{T}^2}
\eeq
and the additional term in the amplitude ${\cal A}$ will be
\beq
\Delta_{2,2}=-\Big(2(eR)_\perp +e_-\frac{(R,p+\tilde{T})_\perp}{R_-}\Big)\frac{Q_2^2}{Q_{1+}\tilde{T}^2}.
\label{del14}
\eeq

{\bf Term $n_1t_2n_3I_{13}$}%$

The diagram is given by
\[I_{13}=-n^-_\beta n^+_\mu n^+_{\alpha_1}n^-_{\alpha_2}\frac{Q_1^2R^2}{p_+Q_{2-}\tilde{T}^2}.\]
Correspondingly, changing $t_2\to q_2$,
\[q_2I_{13}=-n^-_\beta n^+_\mu n^+_{\alpha_1}\frac{Q_1^2R^2}{p_+Q_{2+}\tilde{T}^2}\]
and
\[n_1q_2n_3I_{13}=-n^+_\mu\frac{Q_1^2R^2}{p_+Q_{2+}\tilde{T}^2}.\]
The additional term in the amplitude is
\beq
\Delta_{13}=e_+\frac{Q_1^2R^2}{p_+Q_{2+}\tilde{T}^2}.
\label{del13}
\eeq

{\bf Term $t_1n_2n_3I_{23,2}$}%$

The diagram is given by
\[I_{23,2}=n^-_\beta n^+_\mu n^-_{\alpha_1}n^+_{\alpha_2}\frac{Q_2^2R^2}{p_+Q_{1-}\tilde{T}^2}\ .\]
Correspondingly, with the change $t_1\to q_1$
\[q_1I_{23,2}=n^-_\beta n^+_\mu n^+_{\alpha_2}\frac{Q_2^2R^2}{p_+Q_{1+}\tilde{T}^2}\]
and
\[q_1n_2n_3I_{23.2}=n^+_\mu\frac{Q_2^2R^2}{p_+Q_{1+}\tilde{T}^2}.\]
The additional term in the amplitude is
\beq
\Delta_{23,2}=-e_+\frac{Q_2^2R^2}{p_+Q_{1+}\tilde{T}^2}.
\label{del15}
\eeq

{\bf Total additional terms}

They are six
\beq
\Delta_{tot}=\Delta_{2,1}+\Delta_{3,2}+\Delta_{1,2}+\Delta_{2,2}+\Delta_{13}+\Delta_{23,2}\ .
\eeq
Of these terms $\Delta_{3,2}$, $\Delta_{13}$ and $\Delta_{23,2}$ are proportional
to $e_+$ and vanish in the gauge $e_+=0$. Also all contributions except $\Delta_{2,1}$
are antisymmetric in $Q_1$ and $Q_2$. Because of this they do not appear in the odderon
case.

\section{Vertex RR$\to$RRP}
\subsection{Vertex with Ward identities}
Previous results have a structure, which can naturally be
generalized for larger number of reggeons. For certainty we here study
the case of the vertex RR$\to$RRP.

In our notations with 4 reggeons the vertex as a 4-vector is given by
\beq
\Big(\Gamma_{RR\to RRP}\Big)_\mu=(n_1)_{\alpha_1}(n_2)_{\alpha_2}(n_3)_{\alpha_3}(n_4)_{\alpha_4}
D_{\alpha_1\alpha_2\alpha_3\alpha_4\mu} ,
\label{vertt10}
\eeq
where summation over repeated indices is understood in the Lorentz metric.
For brevity, we rewrite it as
\beq
\Gamma_{RR\to RRP}=n_1n_2n_3n_4D .
\label{vertt1}
\eeq
As before, $D$ is the sum of purely QCD diagrams $D_0$ and diagrams
with induced vertices for reggeons 1,2,3 and 4
\beq
D=D_0+\sum_{i=1}^4I_i+\sum_{i<k=2}^4I_{ik}+\sum_{i<k<l=3}^4I_{ikl}\ ,
\label{a2}
\eeq
where $I_i$, $I_{ik}$ and $I_{ikl}$ are sum of diagrams with induced vertices.
Using (\ref{n}) we have
\beq
\Gamma_{RR\to RRP}=\prod_{i=1}^4(q_i-t_i) D .
\label{vertt2}
\eeq

We find 16 terms. Taking into account that induced vertices are longitudinal
we find these terms as
\[a_1=q_1q_2q_3q_4D,\]
\[a_2=-q_1q_2q_3t_4(D_0+I_1+I_2+I_3+I_{12}+I_{13}+I_{23}+I_{123}),\]
\[a_3=-q_1q_2t_3q_4(D_0+I_1+I_2+I_4+I_{12}+I_{14}+I_{24}+I_{124}),\]
\[a_4=-q_1t_2q_3q_4(D_0+I_1+I_3+I_4+I_{13}+I_{14}+I_{34}+I_{134}),\]
\[a_5=-t_1q_2q_3q_4(D_0+I_2+I_3+I_4+I_{23}+I_{24}+I_{34}+I_{234}),\]
\[a_6=q_1q_2t_3t_4(D_0+I_1+I_2+I_{12}),\]
\[a_7=q_1t_2q_3t_4(D_0+I_1+I_3+I_{13}),\]
\[a_8=q_1t_2t_3q_4(D_0+I_1+I_4+I_{14}),\]
\[a_9=t_1q_2q_3t_4(D_0+I_2+I_3+I_{23}),\]
\[a_{10}=t_1q_2t_3q_4(D_0+I_2+I_4+I_{24}),\]
\[a_{11}=t_1t_2q_3q_4(D_0+I_3+I_4+I_{34}),\]
\[a_{12}=-q_1t_2t_3t_4(D_0+I_1),\]
\[a_{13}=-t_1q_2t_3t_4(D_0+I_2),\]
\[a_{14}=-t_1t_2q_3t_4(D_0+I_3),\]
\[a_{15}=-t_1t_2t_3q_4(D_0+I_4),\]
\[a_{16}=t_1t_2t_3t_4D_0 \ .\]

To calculate these terms we use the WI
\beq
q_1n_2n_3n_4(D_0+I_2+I_3+I_4+I_{23}+I_{24}+I_{34}+I_{234})=0
\label{wwi1}
\eeq
and 3 similar ones obtained by $q_1\to q_2,q_3,q_4$,
\beq
q_1q_2n_3n_4(D_0+I_3+I_4+I_{34})=0
\label{wwi2}
\eeq
and 5 similar ones obtained by $q_1q_2\to q_1q_3,q_1q_4,q_2q_3,q_2q_4,q_3q_4$,
\beq
q_1q_2q_3n_4(D_0+I_4)=0
\label{wwi3}
\eeq
and 3 similar ones obtained by $n_4\to n_1,n_2,n_3$
and finally
\beq
q_1q_2q_3q_4D_0=0 .
\label{wwi4}
\eeq

Our strategy remains the same: using WI we express
contractions of momenta $q_i$ with $D_0$.
We rewrite (\ref{wwi1}) as
\[
q_1(q_2-t_2)(q_3-t_3)(q_4-t_4)(D_0+I_2+I_3+I_4+I_{23}+I_{24}+I_{34}+I_{234})=0
\]
and find (using (\ref{wwi4}))
\[
q_1t_2t_3t_4D_0=q_1q_2q_3q_4(I_2+I_3+I_4+I_{23}+I_{24}+I_{34}+I_{234})\]\[
-q_1q_2q_3t_4(D_0+I_2+I_3+I_{23})-q_1q_2t_3q_4(D_0+I_2+I_4+I_{24})
-q_1t_2q_3q_4(D_0+I_3+I_4+I_{34})
\]
\beq
+q_1q_2t_3t_4(D_0+I_2)+q_1t_2q_3t_4(D_0+I_3)+q_1t_2t_3q_4(D_0+I_4) .
\label{eq1}
\eeq
We put it into $a_{12}$
\[
a_{12}=-q_1t_2t_3t_4I_1-
q_1q_2q_3q_4(I_2+I_3+I_4+I_{23}+I_{24}+I_{34}+I_{234})\]\[
+q_1q_2q_3t_4(I_2+I_3+I_{23})+q_1q_2t_3q_4(I_2+I_4+I_{24})
+q_1t_2q_3q_4(I_3+I_4+I_{34})\]\[
-q_1q_2t_3t_4I_2-q_1t_2q_3t_4I_3-q_1t_2t_3q_4I_4
+(q_1q_2q_3t_4+q_1q_2t_3q_4+q_1t_2q_3q_4\]\beq
-q_1q_2t_3t_4-q_1t_2q_3t_4-q_1t_2t_3q_4)D_0 \ .
\label{eq2}
\eeq
We have also similar expression for $a_{13}$, $a_{14}$ and $a_{15}$.

Now we rewrite (\ref{wwi2}) as
\[
q_1q_2(q_3-t_3)(q_4-t_4)(D_0+I_3+I_4+I_{34})=0 ,
\]
wherefrom we find
\beq
q_1q_2t_3t_4D_0=-q_1q_2q_3q_4(I_3+I_4+I_{34})
+q_1q_2t_3q_4(D_0+I_4)+q_1q_2q_3t_4(D_0+I_3) .
\label{eq3}
\eeq

We rewrite (\ref{wwi3}) as
\[
q_1q_2q_3(q_4-t_4)(D_0+I_4)=0 ,\]
wherefrom
\[
q_1q_2q_3t_4D_0=q_1q_2q_3q_4I_4 \ .
\]
Similarly,
\[
q_1q_2t_3q_4D_0=q_1q_2q_3q_4I_3 \ ,
\]
\[
q_1t_2q_3q_4D_0=q_1q_2q_3q_4I_2 \ ,
\]
\[
t_1q_2q_3q_4D_0=q_1q_2q_3q_4I_1\ .
\]
Using this we find
\[
q_1q_2t_3t_4D_0=
-q_1q_2q_3q_4I_{34}+q_1q_2t_3q_4I_4+q_1q_2q_3t_4I_3 \ .
\]
We put this into $a_6$
\beq
a_6=q_1q_2t_3t_4(I_1+I_2+I_{12})+
q_1q_2t_3q_4I_4+q_1q_2q_3t_4I_3
-q_1q_2q_3q_4 I_{34} \ .
\label{eq4}
\eeq

Other similar formulas are
\[
q_1t_2q_3t_4D_0=
-q_1q_2q_3q_4I_{24}+q_1q_2q_3t_4I_2+q_1t_2q_3q_4I_4
,\]
\[
q_1t_2t_3q_4D_0=
-q_1q_2q_3q_4I_{23}+q_1q_2t_3q_4I_2+q_1t_2q_3q_4I_3
,\]
\[
t_1q_2q_3t_4D_0=
-q_1q_2q_3q_4I_{14}+q_1q_2q_3t_4I_1+t_1q_2q_3q_4I_4
,\]
\[
t_1q_2t_3q_4D_0=
-q_1q_2q_3q_4I_{13}+q_1q_2t_3q_4I_1+t_1q_2q_3q_4I_3
,\]
\[
t_1t_2q_3q_4D_0=
-q_1q_2q_3q_4I_{12}+q_1t_2q_3q_4I_1+t_1q_2q_3q_4I_2
.\]

Using these formulas we find
\[
q_1t_2t_3t_4D_0=-q_1q_2q_3t_4I_{23}-q_1q_2t_3q_4I_{24}-q_1t_2q_3q_4I_{34}\]\[
+q_1q_2t_3t_4I_2+q_1t_2q_3t_4I_3+q_1t_2t_3q_4I_4+q_1q_2q_3q_4I_{234}
,\]
\[
t_1q_2t_3t_4D_0=-q_1q_2q_3t_4I_{13}-q_1q_2t_3q_4I_{14}-t_1q_2q_3q_4I_{34}\]\[
+q_1q_2t_3t_4I_1+t_1q_2q_3t_4I_3+t_1q_2t_3q_4I_4+q_1q_2q_3q_4I_{134}
,\]
\[
t_1t_2q_3t_4D_0=-q_1q_2q_3t_4I_{12}-q_1t_2q_3q_4I_{14}-t_1q_2q_3q_4I_{24}\]\[
+q_1t_2q_3t_4I_1+t_1q_2q_3t_4I_2+t_1t_2q_3q_4I_4+q_1q_2q_3q_4I_{124}
,\]
\[
t_1t_2t_3q_4D_0=-q_1q_2t_3q_4I_{12}-q_1t_2q_3q_4I_{13}-t_1q_2q_3q_4I_{23}\]\[
+q_1t_2t_3q_4I_1+t_1q_2t_3q_4I_2+t_1t_2q_3q_4I_3+q_1q_2q_3q_4I_{123}
.\]

This allows to exclude all terms with $D_0$ multiplied by $q_i$.
Call $-X_3$ the sum of such terms in $a_{12}+a_{13}+a_{14}+a_{15}$. We find
\[X_3=q_1q_2q_3t_4(I_{23}+I_{13}+I_{12})+q_1q_2t_3q_4(I_{24}+I_{14}+I_{12})\]\[+
q_1t_2q_3q_4(I_{34}+I_{14}+I_{13})+t_1q_2q_3q_4(I_{34}+I_{24}+I_{23})\]\[
-q_1q_2t_3t_4(I_1+I_2)-q_1t_2q_3t_4(I_1+I_3)-q_1t_2t_3q_4(I_1+I_4)\]\[
-t_1q_2q_3t_4(I_2+I_3)-t_1q_2t_3q_4(I_2+I_4)-t_1t_2q_3q_4(I_3+I_4)\]\[
-q_1q_2q_3q_4(I_{123}+I_{134}+I_{124}+I_{234}) .
\]
Call $X_2$ the similar sum of terms in $a_6+a_7+a_8+a_9+a_{10}+a_{11}$. We find
\[X_2=-q_1q_2q_3q_4(I_{12}+I_{13}+I_{14}+I_{23}+I_{24}+I_{34})
+q_1q_2t_3q_4(I_1+I_2+I_4)+q_1q_2q_3t_4(I_1+I_2+I_3)\]\[+
q_1t_2q_3q_4(I_1+I_3+I_4)+t_1q_2q_3q_4(I_2+I_3+I_4) .\]
Call finally $X_1$ the similar sum of terms in $a_2+a_3+a_4+a_5$. We find
\[X_1=-q_1q_2q_3q_4(I_1+I_2+I_3+I_4) .\]

Summing $X_1+X_2-X_3$ we find that all terms with the product $q_1q_2q_3q_4$ cancel term $a_1$
and the rest terms give in the sum
\[q_1q_2q_3t_4(I_3+I_2+I_1+I_{23}+I_{13}+I_{12})+
q_1q_2t_3q_4(I_1+I_2+I_4+I_{13}+I_{24}+I_{12})\]\[+
q_1t_2q_3q_4(I_3+I_4+I_1+I_{14}+I_{34}+I_{13})+
t_1q_2q_3q_4(I_2+I_3+I_4+I_{23}+I_{24}+I_{34})\]\[
-q_1q_2t_3t_4(I_1+I_2)-q_1t_2q_3t_4(I_1+I_3)-q_1t_2t_3q_4(I_1+I_4)\]\[
-t_1q_2q_3t_4(I_2+I_3)-t_1q_2t_3q_4(I_2+I_4)-t_1t_2q_3q_4(I_3+I_4) .\]

These terms are to be summed with those in $a_2$--$a_{15}$ which do not contain $D_0$.
In the sum we get
\[
\Gamma_{RR\to RRP}=t_1t_2t_3t_4D_0
-q_1q_2q_3t_4I_{123}-q_1q_2t_3q_4I_{124}-q_1t_2q_3q_4I_{134}-t_1q_2q_3q_4I_{234}\]\[
+q_1q_2t_3t_4I_{12}+q_1t_2q_3t_4I_{13}+q_1t_2t_3q_4I_{14}
+t_1q_2q_3t_4I_{23}+t_1q_2t_3q_4I_{24}+t_1t_2q_3q_4I_{34}\]\beq
-q_1t_2t_3t_4I_1-t_1q_2t_3t_4I_2-t_1t_2q_3t_4I_3-t_1t_2t_3q_4I_4 \ .
\label{vertt3}
\eeq
As in the previous section, all indefinite expressions
like $(tn^-)/Q_-$ at $Q_-=0$ are to be replaced by
$(qn^-)/Q_-$.

\subsection{Diagrams}
\begin{figure}
\hspace*{1 cm}
\epsfig{file=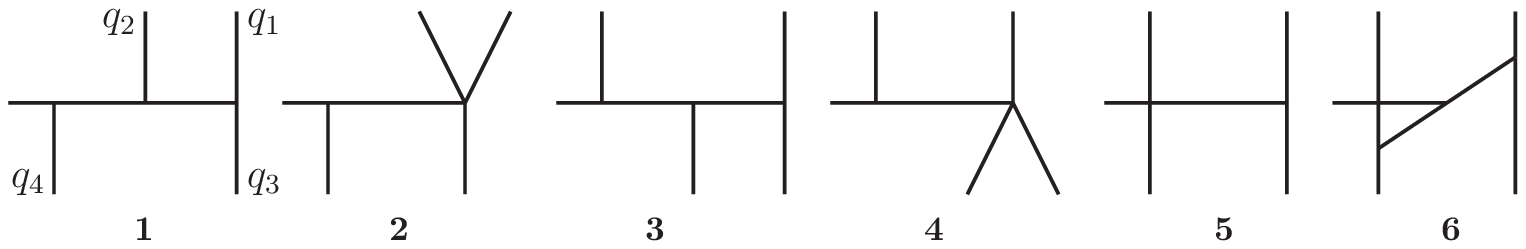, width=14 cm}
\vskip 0.4cm
\hspace*{1.5 cm}
\epsfig{file=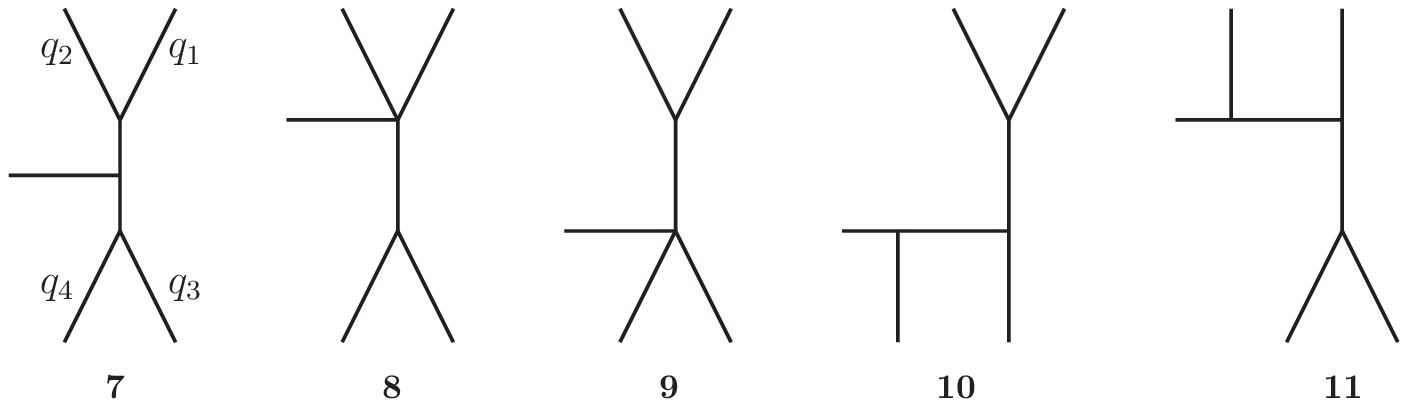, width=13 cm}
\caption{Diagrams for the vertex RR$\to$RRP with QCD vertices.
Simple external lines directed upwards and downwards actually refer to reggeons.}
\label{fig4}
\end{figure}

\begin{figure}
\hspace*{1 cm}
\epsfig{file=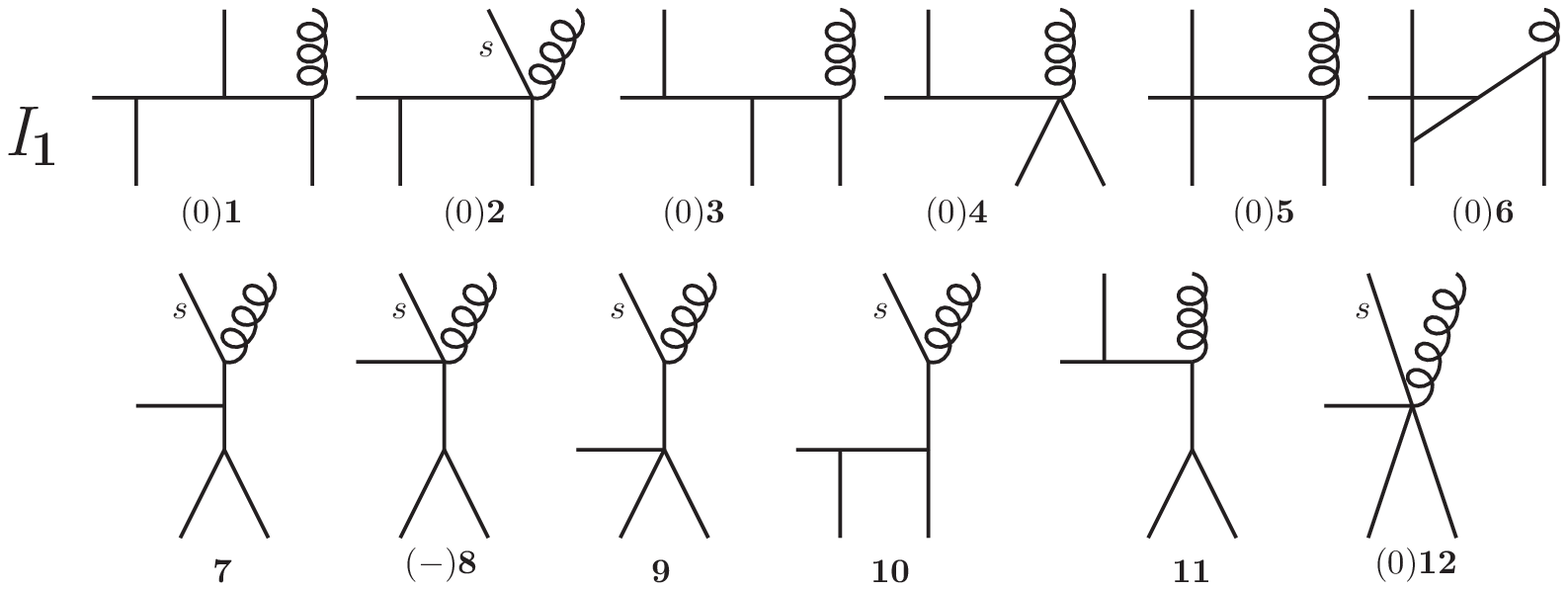, width=14 cm}
\vskip 0.75cm
\hspace*{1 cm}
\epsfig{file=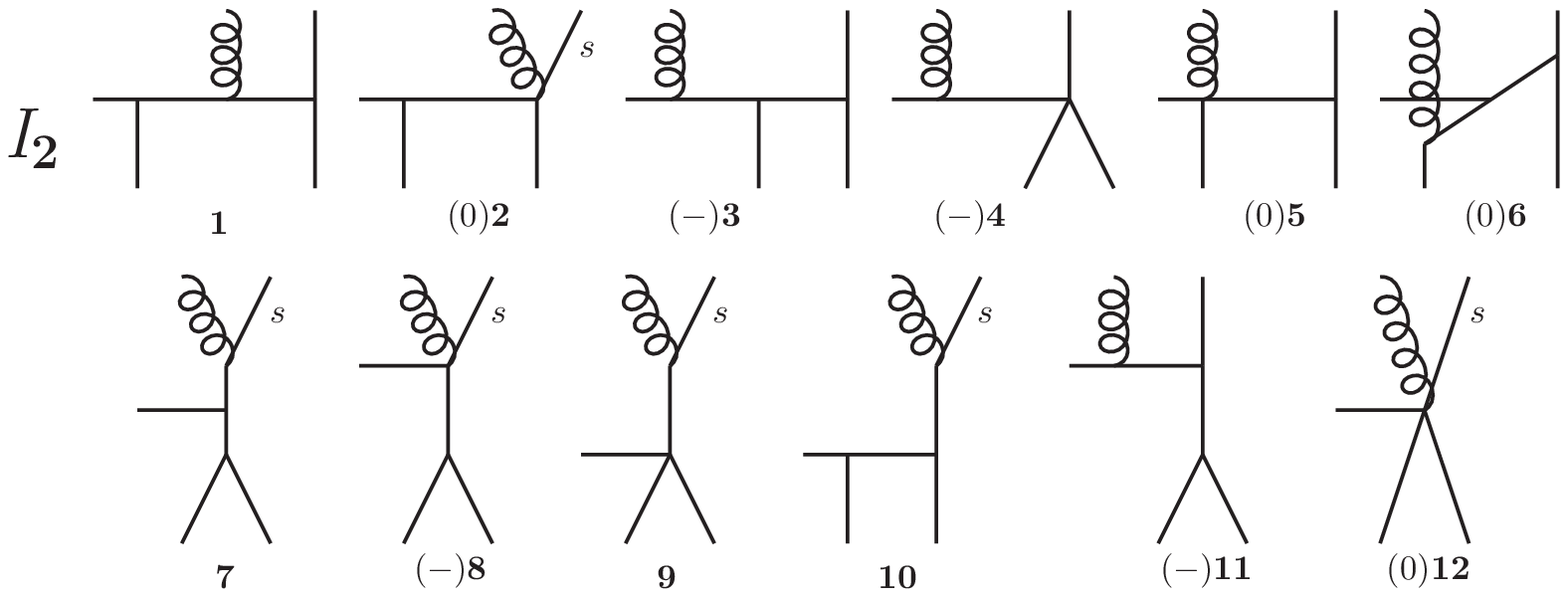, width=14 cm}
\vskip 0.75cm
\hspace*{1 cm}
\epsfig{file=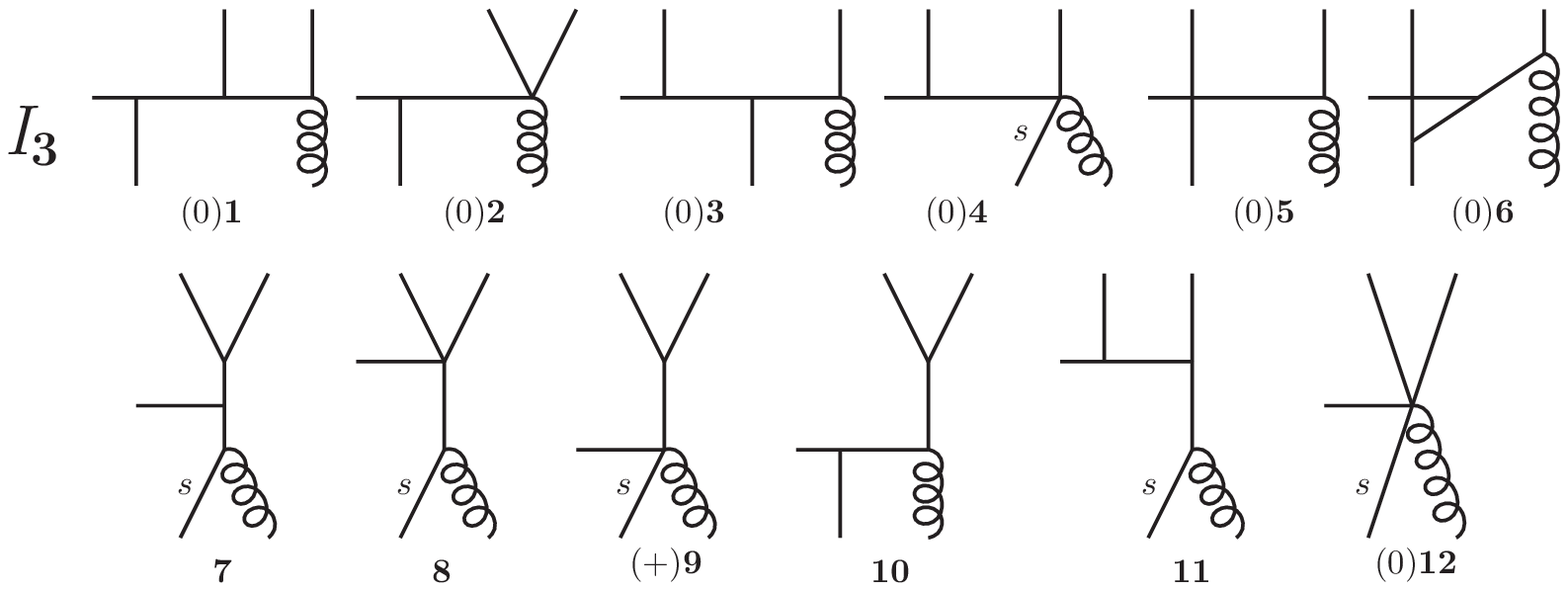, width=14 cm}
\vskip 0.75cm
\hspace*{1 cm}
\epsfig{file=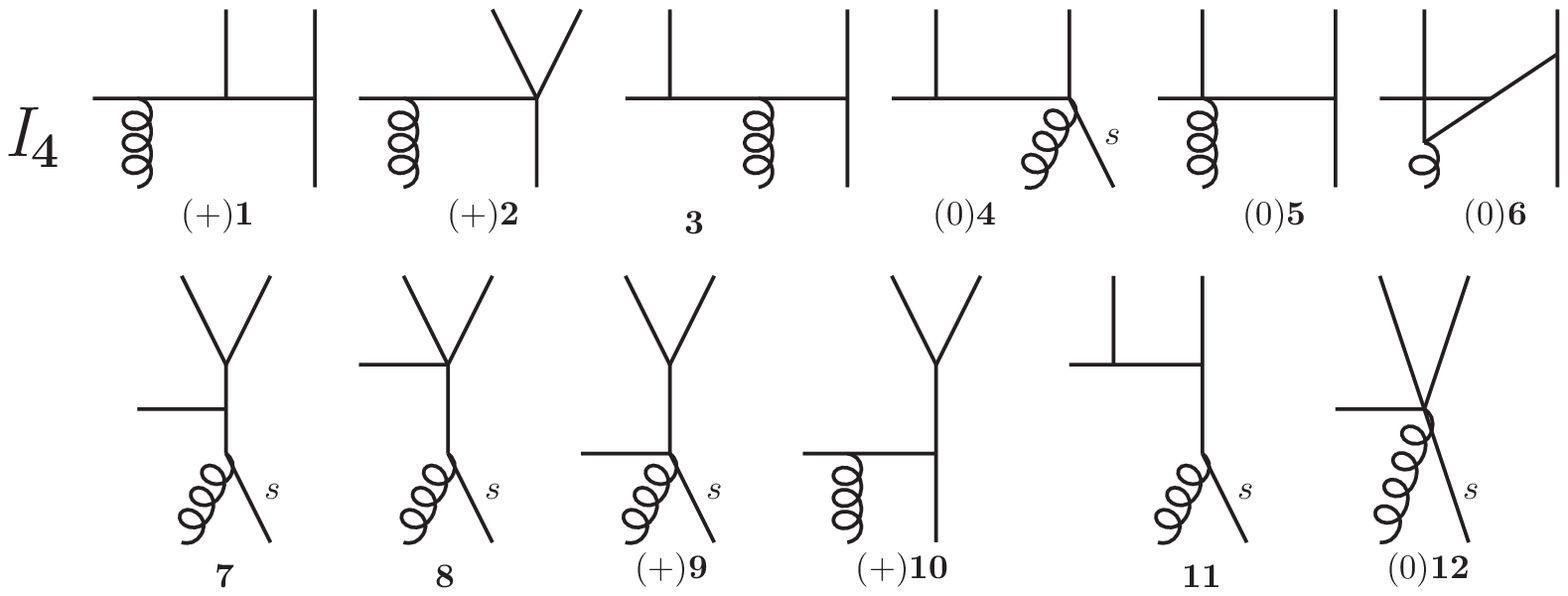, width=14 cm}
\caption{Diagrams for the vertex RR$\to$RRP and Ward identities
with one induced vertex.
Simple external lines directed upwards and downwards actually refer to reggeons.}
\label{fig5}
\end{figure}

All relevant diagrams are shown in Figs. \ref{fig4}-\ref{fig7}.
The QCD diagrams which correspond to term $D_0$ are shown in Fig. \ref{fig4}.
Diagrams with one induced vertex $I_i$, $i=1,2,3,4$ are presented in Fig. \ref{fig5}.
Those with two induced vertices $I_{ik}$, $i<k$ are shown in Fig. \ref{fig6}.
Finally, Fig. \ref{fig7} shows diagrams $I_{ikl}$, $i<k<l$ with three induced vertices.

\begin{figure}
\hspace*{1 cm}
\epsfig{file=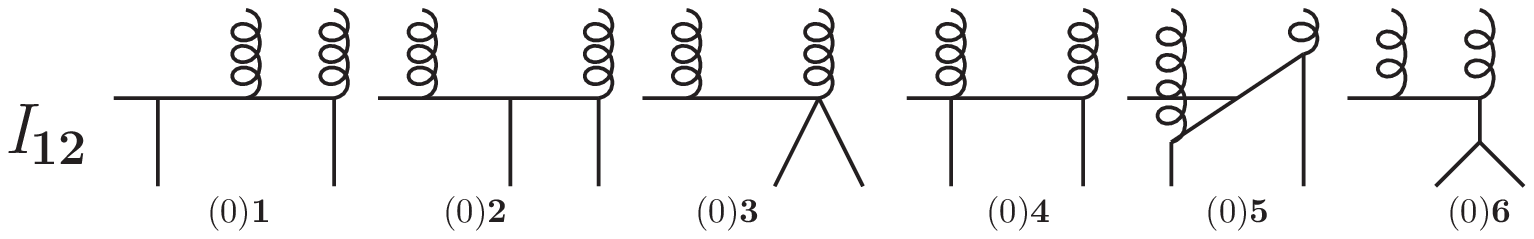, width=14 cm}
\vskip 0.75cm
\hspace*{2 cm}
\epsfig{file=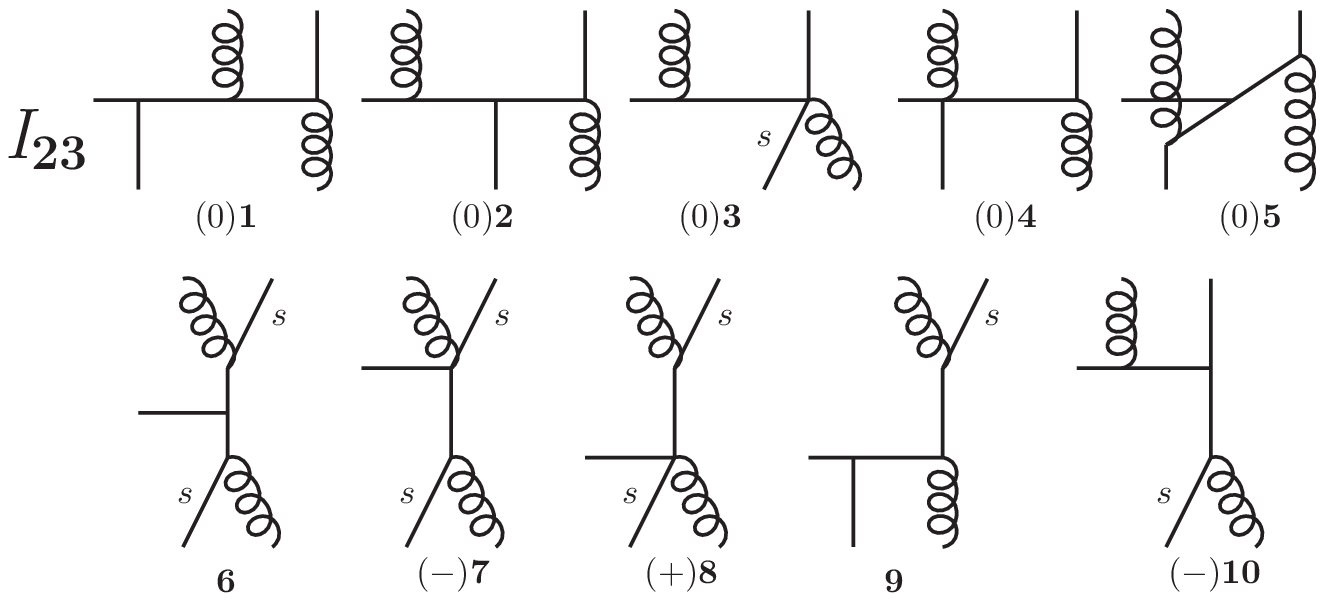, width=12 cm}
\vskip 0.75cm
\hspace*{1 cm}
\epsfig{file=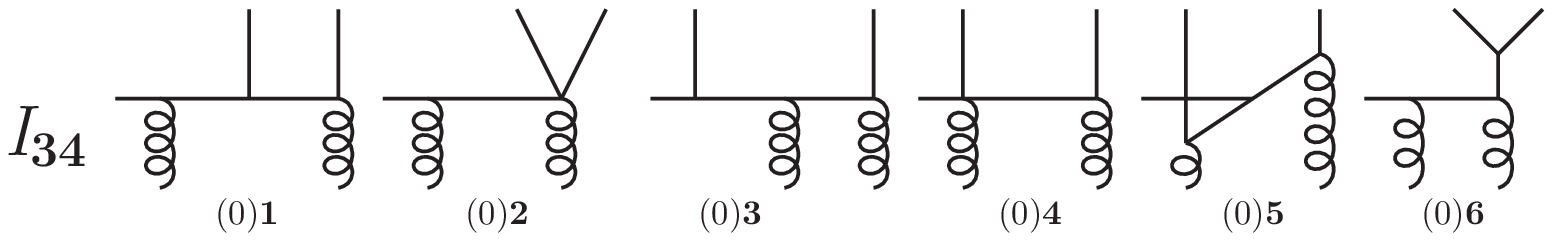, width=14 cm}
\vskip 0.75cm
\hspace*{2 cm}
\epsfig{file=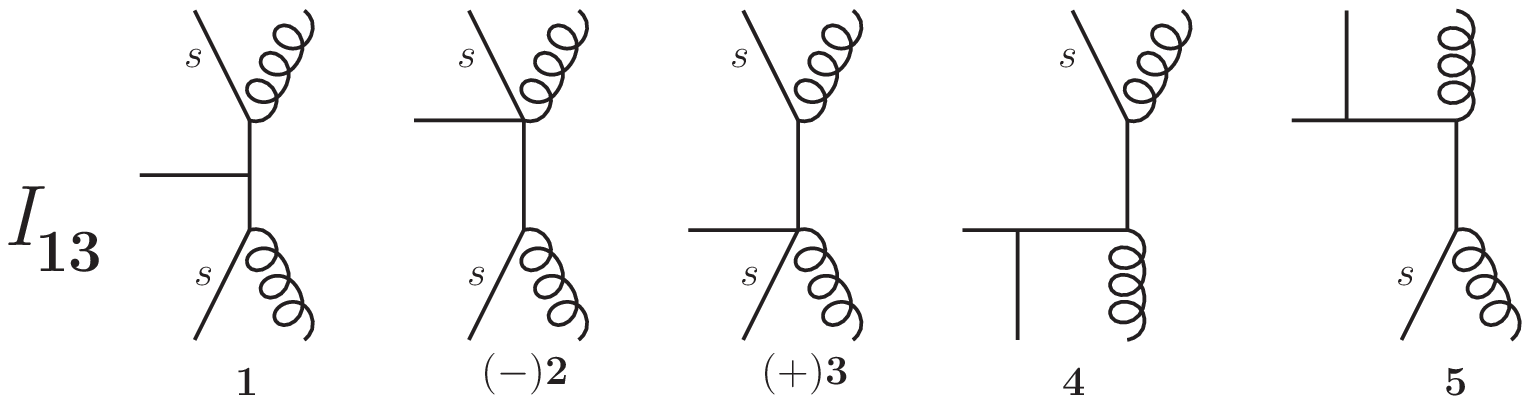, width=12 cm}
\vskip 0.75cm
\hspace*{2 cm}
\epsfig{file=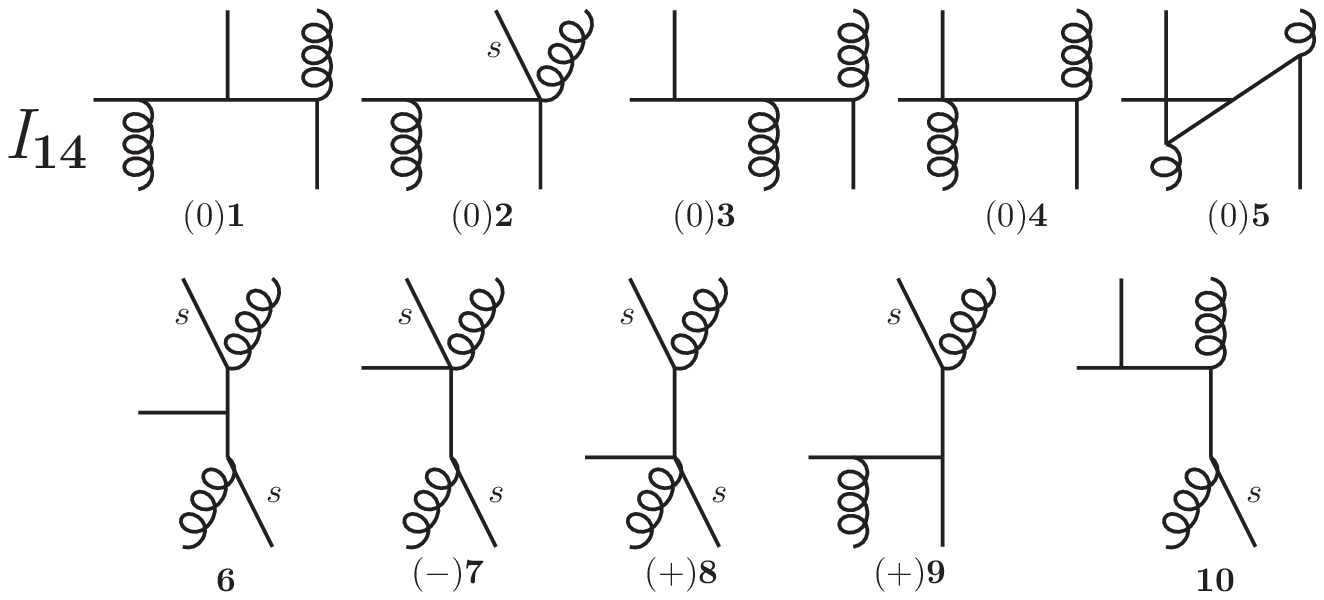, width=12 cm}
\vskip 0.75cm
\caption{Diagrams for the vertex RR$\to$RRP and Ward identities
with two induced vertices.
Simple external lines directed upwards and downwards actually refer to reggeons.}
\end{figure}

\begin{figure}
\addtocounter{figure}{-1}
\hspace*{1.5 cm}
\epsfig{file=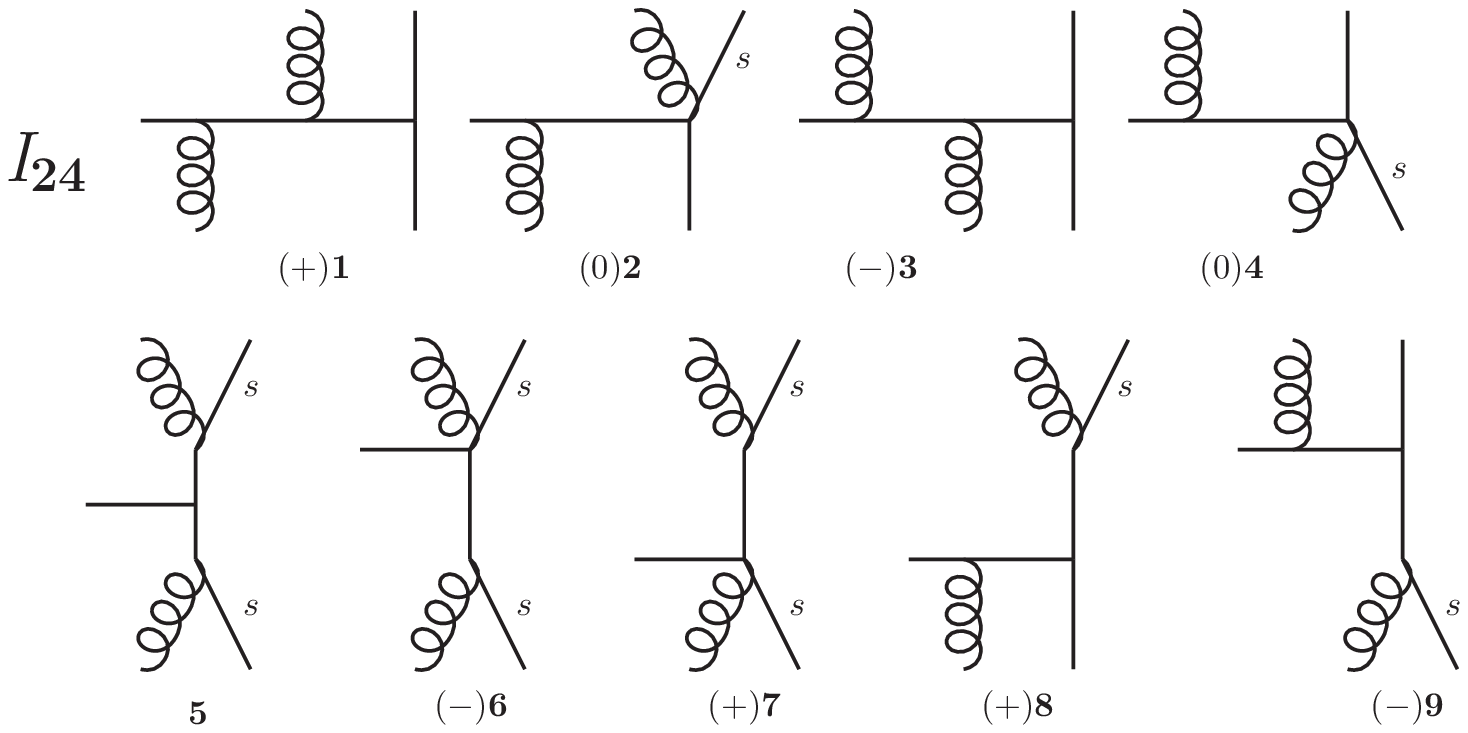, width=13 cm}
\caption{Diagrams for the vertex RR$\to$RRP and Ward identities
with two induced vertices (continuation).}
\label{fig6}
\end{figure}

\begin{figure}[h]
\hspace*{1.5 cm}
\epsfig{file=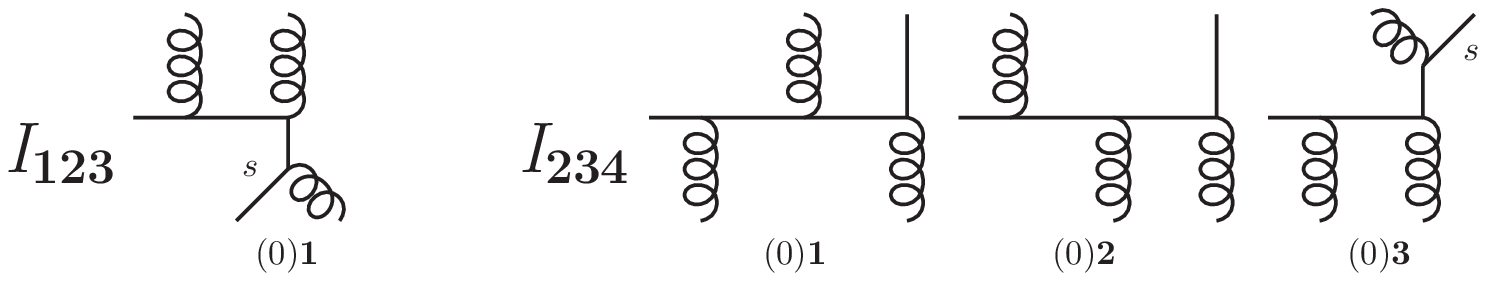, width=12.5 cm}
\vskip 0.75cm
\hspace*{1.5 cm}
\epsfig{file=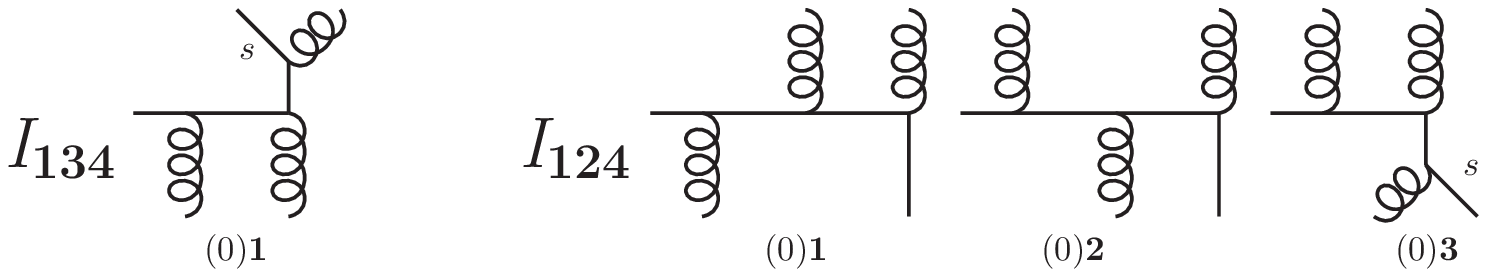, width=12.5 cm}
\caption{Diagrams for the vertex RR$\to$RRP and Ward identities
with three induced vertices.
Simple external lines directed upwards and downwards actually refer to reggeons.}
\label{fig7}
\end{figure}

Note that from the 11 QCD diagrams shown in Fig. \ref{fig4} only 6 in the
upper part give contribution to the vertex in the original expression (\ref{vertt1}).
Those in the lower line give zero due to the property of the 3-gluon vertex
with two of its legs with reggeon momenta.
Likewise from the total 102 induced diagrams
only 36 take part in the original expression (\ref{vertt1}).
From the diagrams with a single induced vertex they
are $I_{i,1},I_{i,3},I_{i,4}, I_{i,5}, I_{i,6}$ for $i=1,2$
and $I_{i,1},I_{i,2},I_{i,3}, I_{i,5}, I_{i,6}$ for $i=3,4$ (20 diagrams in all).
From the diagrams with two induced vertices they are
\[I_{12,2}, I_{12,5}, I_{23,1}, I_{23,2},I_{23,4},I_{23,5}, I_{34,1}, I_{34,5},\]
\[I_{14,1}, I_{14,3},I_{14,4}, I_{14,5}, I_{24,1}, I_{24,3}\]
(14 diagrams in all).
And from the diagrams with three induced vertices two diagrams $I_{234,1}$ and $I_{124,2}$.
All the rest give zero after multiplication by $n_1n_2n_3n_4$. So the total number of non-zero diagrams
in the original expression (\ref{vertt1}) is 42.

Expression (\ref{vertt3}) obtained after application of the Ward identities contains
all 11  QCD diagrams and  additional contributions coming from the induced diagrams.
%Not all induced diagrams give non-zero contributions.
Many of them give zero contributions
to (\ref{vertt3}). Similarly to Fig. \ref{fig2} they are marked with index (0) in Figs. \ref{fig5}-\ref{fig7}.
The total number of induced diagrams which give non-zero contributions turns out to be 48.
So the use of Ward identities leads to 59 non-zero diagrams, which is 17 diagrams more than the original expression.
Still this number can be diminished after the choice of one of the gauges with either $e_+=0$ or $e_-=0$.
As in Fig. \ref{fig2} we mark by indices $(\pm)$ those diagrams in which the leg corresponding to
the emitted real gluon carries factor $n^{\pm}$. Clearly diagrams marked with $(\pm)$ will vanish in the gauge
$e_{\pm}=0$. The number of such  diagrams with a fixed sort of $n$ is 12.
This reduces the number of additional diagrams to 36 and the total number to 47.

We are not going to write out the numerous additional contributions to
(\ref{vertt3}) coming from the induced diagrams here. They are too many and their derivation is straightforward and
clear from the similar derivation for the vertex RR$\to$RP in the previous section. We only want to mention, that
although the whole set of induced diagrams apart from the vertices $V_1,...,V_4$, Eq. (\ref{rules}) contains the
vertex for transition of a reggeon in 4 gluons, the diagrams with this latter vertex give no contribution
to the final expression (\ref{vertt3}). So we do not need the explicit form of this vertex
(which can easily be found using the
recurrent relation in ~\cite{antonov}).

The obtained contributions have naturally to be symmetrized in ingoing reggeons (with momenta $q_1$ and $q_2$)
and in outgoing reggeons ($q_3$ and $q_4$).

\section{Conclusions}
Following the suggestion in ~\cite{BLV} we studied Ward identities for
effective theory of interacting gluons and
reggeons in general configurations for vertices RR$\to$RP and RR$\to$RRP.
For these Ward identities we have found that though the set of initial
reggeon diagrams is indeed to be reduced to exclude transition of a
particular reggeon in a single gluon, many more new diagrams are to be
included corresponding to the replacement of this reggeon by the gluon.
We have found the final expressions which express the vertices via the
QCD diagrams convoluted with the transverse momenta plus a certain
number of additional diagrams, coming from the induced ones. It turns
out that among these additional diagrams there appear singular ones
poorly defined in the Regge kinematics. We formulated the rule to fix
their definition.

In the end in the general gauge the final number of diagrams in the
expression obtained by means of Ward identities turns out to be larger
than in the original expression: 59 against 42 for the vertex RR$\to$RRP.
Use of special gauges allows to reduce the former number to 47: still 5 diagrams
more than in the original expression. The number of additional diagrams
may be further reduced in special colour configurations, such as the one
for the odderon, considered in \cite{BFLV}.

Note also that presence of singular diagrams casts doubts on the
possibility to get better convergence in the longitudinal momenta
applying Ward identities, since in the additional terms which come
from them transverse momenta are to be replaced by the total ones.

\section{Acknowledgements}
The authors acknowledge Saint-Petersburg State University
for a research grant 11.38.223.2015. This study has been also
supported by the RFBR grant 15-02-02097.

\end{document}